\documentclass[12pt,twoside,a4paper,final]{article}
\usepackage[leqno]{amsmath}
\usepackage{amsbsy}
\usepackage{fullpage}
\usepackage{xspace}
\usepackage{comment}
\usepackage{graphicx}
\usepackage{natbib}
\usepackage{appendix}
\usepackage{lscape}

\usepackage{bm}
\usepackage{rotating}
\usepackage{rotfloat}
\usepackage{authblk}

\newcommand{\XXi}{\boldsymbol{\Xi}}
\newcommand{\llambda}{\boldsymbol{\lambda}}
\newcommand{\LLambda}{\boldsymbol{\Lambda}}
\newcommand{\eeta}{\boldsymbol{\eta}}
\newcommand{\btheta}{\boldsymbol{\theta}}
\newcommand{\btau}{\boldsymbol{\tau}}
\newcommand{\bbeta}{\boldsymbol{\beta}}
\newcommand{\bgamma}{\boldsymbol{\gamma}}

\newcommand{\OOmega}{\boldsymbol{\Omega}}
\newcommand{\aalpha}{\boldsymbol{\alpha}}
\newcommand{\PPi}{\boldsymbol{\Pi}}

\usepackage{moreverb}
\usepackage{bm}

\newcommand\BibTeX{{\rmfamily B\kern-.05em \textsc{i\kern-.025em b}\kern-.08em
T\kern-.1667em\lower.7ex\hbox{E}\kern-.125emX}}

\begin{document}

\title{\textbf{Vertical modeling: analysis of competing risks data with a cure proportion}}



\author{M. A. Nicolaie$^{a}$ \thanks{Corresponding author. E-mail: mioara.nicolaie@uclouvain.be \vspace{6pt}} \ and J.M.G. Taylor$^{b}$ and C. Legrand$^{a}$\\
       \vspace{6pt} $^{a}${\em{\small{\fontsize{8pt}{8pt} Institute of Statistics, Biostatistics and Actuarial Sciences, Catholic University of Louvain, Voie du Roman Pays 20, bte L1.04.01, 1348 Louvain-la-Neuve, Belgium}}} \\ $^{b}${\em{\small{\fontsize{8pt}{8pt} School of Public Health, University of Michigan, M4509 SPH II, 1415 Washington Heights, Ann Arbor, Michigan 48109-2029, USA}}}}

\maketitle


\date{}



\begin{abstract}

In this paper, we extend the vertical modeling approach for the analysis of survival data with competing risks to incorporate a cured fraction in the population, that is, a proportion of the population for which none of the competing events can occur. The proposed method has three components: the proportion of cure, the risk of failure, irrespective of the cause, and the relative risk of a certain cause of failure, given a failure occurred. Covariates may affect each of these components. An appealing aspect of the method is that it is a natural extension to competing risks of the semi-parametric mixture cure model in ordinary survival analysis; thus, causes of failure are assigned only if a failure occurs. This contrasts with the existing mixture cure model for competing risks of Larson and Dinse, which conditions at the onset on the future status presumably attained. Regression parameter estimates are obtained using an EM-algorithm. The performance of the estimators is evaluated in a simulation study. The method is illustrated using a melanoma cancer data set.

\end{abstract}



\section{Introduction}\label{intro}

In medicine, the risk of failure from a given disease or the chance of getting cured of it are of interest to the diseased patients as well as to the treating physicians; this helps patients at risk to make important life decisions and it helps physicians in treatment selection. However, investigating this type of information in a clinical study involving time-to-event data requires accounting for multiple possible outcomes: either failure, due to the disease or due to some other cause, or cure. From the statistical point of view, the analysis of several types of failures is incorporated in the competing risks framework. These methods usually assume that all patients will eventually experience one of the possible types of failure if there is sufficient follow-up and, therefore, do not accomodate cure. The presence of cure is suggested when the data includes a considerable number of long-term event-free survivors (censored patients with long follow-up times). Therefore, extending modeling of competing risks to accommodate a cured proportion is an important  issue in understanding this type of data.

In the analysis of single outcome survival data with a cured fraction, cure models address the problem of cure rate estimation, as well as the estimation of the probability of failure due to the disease of interest. They have received a lot of attention both in terms of methodological developments~\cite{Fa:86,SyTay:00,LiTay:02,Yu:04,Ki:13} and applications~\cite{AndDick:11,An:12}. These cure models are mixture models which specify a conditional model for the survival component, given that failure may occur, and a marginal distribution of the binary indicator for whether or not cure can occur. They are formulated either in a parametric or in a semi-parametric way~(\cite{Ku:92,Tay:95,SyTay:00,Peng:00,Peng:03,Co:09}). They share the common feature of allowing the cure rate to be determined from the onset, but to be observed only later in the course of the follow-up, leading to the presence of a sub-population of event-free survivors beyond sufficiently long follow-up.

Analysis of competing risks data with a cure fraction is not well developed. Several authors caste the competing risks model of ~\cite{LasDin:85} into the cure framework. These models express the mixing joint distribution of the time-to-event and type of event variables as the sum of the marginal cause-specific distributions multiplied by the associated mixing proportions. This approach is formulated under the strong, unverifiable assumption that the mixing proportions for failure types and cure indicator is determined but unobserved at the onset. In terms of formulas, this amounts to the following decomposition of the joint distribution of time of failure $T$ and failure type $D$:
\begin{equation}
 P(T,D) = P(T|D)\cdot P(D).
\end{equation}
Examples include the approach of ~\cite{Ch:98}, which imputes the cure indicator for censored patients and uses a Gibbs sampling algorithm for estimation; ~\cite{Ng:98} propose a parametric version, to be used when there are only few failures from the competing causes and ~\cite{Cho:02} discuss extensively a class of multivariate parametric models. This approach could be adopted in contexts where the interest is in assessing the parameters of the conditional failure time distribution given failure type or in the mixing distribution of the different competing failure types.

However, from the inference and interpretation points of view, the case where the model parameters translate directly into natural observable quantities in competing risks is appealing. In this paper, we adopt this perspective and introduce a semi-parametric approach for the analysis of competing risks data with a cure fraction. This approach extends the idea of vertical modeling formulated earlier by ~\cite{Nic:10} in the competing risks framework, and it is based on the following decomposition of the joint distribution of time of failure $T$ and failure type $D$:
\begin{equation}
 P(T,D) = P(D|T)\cdot P(T).
\end{equation}
In the remainder of the article, we demonstrate how this can be applied to the analysis of mixture cure data with several competing causes of failure in the presence of right-censored data. In Section~\ref{sec:VMCF} we introduce in detail our approach. Simulation studies are presented in  Section~\ref{sec:sim}. In Section~\ref{sec:data_an} we illustrate our methods of analysis on a clinical study on melanoma cancer. Section~\ref{sec:disc} concludes with some points for discussion.

\section{Vertical modeling with a cured fraction}\label{sec:VMCF}

\subsection{Notation}

Suppose that data are available from $n$ individuals each of whom can experience one of $J$ terminal, competing events during the period under study or can be subjected to noninformative right censoring. Assume that a non-negligible proportion of individuals does not experience any of the terminal events by the end of the follow-up and assume further that follow-up is long enough to be able to consider individuals with full follow-up as cured from the disease of interest on the basis of some clinical evidence. Such population can naturally be regarded as a mixture population in which two categories of individuals are combined: susceptible (the individual experiences failure, irrespective of the cause) and non-susceptible or cured (the individual is immune to all causes of failure). Let $\tilde{T}$ denote the time-to-event variable, $C$ the right-censoring time variable, and $D$ the terminal event type, where $D \in \{1,\ldots,J\}$. Assign $D = 1$ for the cause of interest and $D \geq 2$ for the competing causes. Let $\textbf{Z}$ denote an $l$-vector of covariates measured at baseline.

Consider a binary random variable $Y$ such that $Y=1$ corresponds to a susceptible individual and $Y=0$ corresponds to a nonsusceptible/cured individual. Note that in this type of study $Y$ is partially observed; it equals $1$ in the case of an event, and it is unobserved in the case of right-censoring. If the latter occurs, the individual has no event observed during the study period, but either the event will eventually take place (the individual is censored and susceptible) or the event will never take place in the future (the individual is censored and non-susceptible).

The observed data for an individual $i$ is $\mathcal{O}_i = (T_i, \Delta_i, Z_i)$, where $T_i = \min(\tilde{T}_i, C_i)$ is the earliest of time-to-event and censoring time, and $\Delta_i = \textbf{1}\{\tilde{T}_i< C_i\} D_i$ is the type of terminal event in the case a terminal event occurs and $0$ in the case of censoring, for $i=1,\ldots,n$. Data from different individuals are assumed to be independent. Assume that $(T,D)$ and $C$ are independent given $\textbf{Z}$.

\subsection{Model formulation}

Our goal is to develop and implement an approach to determine whether a terminal event would occur (whether an individual is susceptible, which hereafter is referred to as \emph{incidence}) and, conditional on being susceptible, when the event might occur and which type of event it might be, given a failure occurred (which hereafter are jointly referred to as \emph{latency}). In addition, we are interested in assessing covariate effects on incidence and latency.

The incidence part is completely specified by the probability distribution $P(Y)$. Denote $P(Y = 1) = p$; thus, $1-p$ represents the proportion of individuals who get cured. For the latency part, we aim to extend the vertical modeling approach, earlier proposed by ~\cite{Nic:10}, to the mixture cure model framework. We shall refer to this competing risks mixture cure model as vertical modeling with a cured fraction (VMCF).

The main idea behind the modeling of the latency part of VMCF is to specify the conditional (on $Y=1$) joint distribution $P(T,D|Y=1)$ as
\begin{equation}\label{eq:decomp-vm}
     P(T,D|Y=1) = P(T|Y=1)\cdot P(D|T,Y=1),
\end{equation}
that is, the product of the conditional (on $Y=1$) failure rate $P(T|Y=1)$ and of the conditional distribution of the causes of failure, given a failure occurred $P(D|T,Y=1)$. If we assume that the survival time $T$ is continuous, we define the conditional (on $Y=1$) total hazard by
\begin{equation}\label{eq:total_haz}
    \lambda_{\bullet}(t|Y=1) = \lim_{\Delta t \rightarrow 0}\frac{P( t\leq T\leq t+\Delta t| T \geq t, Y = 1)}{\Delta t}
\end{equation}
and its cumulative counterpart by $\Lambda_{\bullet}(t|Y=1) = \int_{0}^{t}\lambda_{\bullet}(u|Y=1) du$. The former specifies the conditional (on $Y=1$) failure distribution $P(T|Y=1)$. In turn, the conditional (on $Y=1$) survival function of susceptible individuals, defined as $S(t|Y=1) = P(T>t|Y=1)$ is given by
\begin{equation}\label{eq:cond_surv}
  S(t|Y=1) = \exp(-\Lambda_{\bullet}(t|Y=1)).
\end{equation}
We emphasize that $S(t|Y=1)$ is a proper survival function in the sense that $\lim_{t\rightarrow \infty}S(t|Y=1) = 0$. Note that the conditional (on $Y=0$) survival function of non-susceptible is degenerate, that is, $P(T>t|Y=0)=1$. We define the conditional (on $Y=1$) relative cause-specific hazard of cause $j$ at time $t$ by
\begin{equation}\label{eq:rel_haz}
  \pi_{j}(t|Y=1) = P(D = j|\widetilde{T} = t, Y= 1),\qquad j = 1,\ldots,J.
\end{equation}
Note that the probability $\pi_j(t| \widetilde{T} = t, Y = 1)$ deals with failure time and cause, therefore its estimation involves only the (observed) susceptible individuals. This implies that (\ref{eq:rel_haz}) can be expressed as:
\begin{equation}\label{eq:rel_haz_rewri}
  P(D = j|\widetilde{T} = t, Y= 1) = P(D = j|\widetilde{T} = t),\qquad j = 1,\ldots,J.
\end{equation}
As a consequence, we suppress the dependence on $Y=1$ of $\pi_{j}(t|Y=1)$ and, in the following, we will denote it simply by $\pi_{j}(t)$. Thus the vector $(\pi_{j}(t))_{j=1,\ldots,J}$ denotes the conditional (on $\widetilde{T}=t$) distribution $P(D |\widetilde{T} = t)$, with $\Sigma_j \pi_j(t) = 1$ for all $t$.
Thus the vector $(\lambda_{\bullet}(t|Y=1), (\pi_{j}(t))_{j=1,\ldots,J})$ completely specifies the latency part of VMCF.

The conditional (on $Y=1$) cumulative incidence function of cause $j$, defined as $F_j(t|Y=1) = P(T \leq t, D = j|Y = 1)$, can be obtained as
\begin{equation}\label{eq:cond_cif}
  F_j(t|Y=1) = \int_{0}^{t}\lambda_{\bullet}(u|Y=1)\pi_{j}(u)S(u-|Y=1)du,\qquad j = 1,\ldots,J.
\end{equation}
The marginal survival function, which is defined as $S_{pop}(t) = P(T\geq t)$ can be expressed as
\begin{eqnarray}\label{eq:pop_surv}
  S_{pop}(t) &=& P(Y=1)P(T\geq t|Y=1) + P(Y=0)P(T\geq t|Y=0)\\
             &=& p \cdot S(t|Y=1) + 1 - p.\notag
\end{eqnarray}
Note that $S_{pop}(t)$ is an improper survival function in the sense that $\lim_{t\rightarrow \infty}S_{pop}(t) = 1-p$.

\subsection{Specific models}

For the incidence $p(\textbf{X}) = P(Y = 1| \textbf{X})$, where $\textbf{X} \subseteq \textbf{Z}$, we postulate the usual binary regression models:
\begin{equation}\label{eq:transf-models}
 g(p(\textbf{X})) = \mbox{\boldmath${\beta}$}^{\top} \textbf{X}^{*},
\end{equation}
where $\textbf{X}^{*} = (1, \textbf{X})$, \mbox{\boldmath${\beta}$} stands for a vector of unknown regression parameters, including an intercept, and $g$ is a known differentiable link function. This includes the logistic link model $g(p) = \log\frac{p}{1-p}$, the complementary log-log link $g(p) = \log(-\log(1-p))$ and the probit link $g(p) = \Phi^{-1}(p)$, where $\Phi^{-1}$ is the distribution function of a standard normal distribution.

For the conditional (on $Y=1$) total hazard we postulate a Cox proportional hazards model:
\begin{equation}\label{eq:cond_haz_pop_model}
    \lambda_{\bullet}(t|Y=1, \textbf{Z}) = \lambda_{0}(t|Y=1)\exp{(\bgamma^{\top} \textbf{Z})},
\end{equation} where $\lambda_{0}(\cdot|Y = 1)$ is an unspecified conditional (on $Y=1$) baseline hazard and $\mathbf{\gamma}$ stands for a vector of unknown regression parameters. Let $t_0=0$ and denote by $t_{1}\leq t_{2}\leq \ldots \leq t_{{K}}$ the $K$ ordered event times. Assume that $\lambda_{0}(t|Y=1)$ is a step function such that
\begin{equation*}
    \lambda_{0}(t|Y=1) = \lambda_{0}(t_{l}|Y=1), \qquad \textrm{all}\ t \in [t_{l}, t_{l+1})
\end{equation*}
for $l \in \{0,\ldots, K-1\}$.

For the relative hazards we specify
\begin{equation}\label{eq:rel_haz_model}
     \pi_j(t| \textbf{U}) = \frac{\exp(\mathbf{\kappa}_j^{\top} \textbf{B}(t) + \mathbf{\upsilon}_j^{\top} \textbf{U})}{\sum_{l=1}^{J}\exp(\mathbf{\kappa}_l^{\top} \textbf{B}(t) + \mathbf{\upsilon}_l^{\top} \textbf{U})},\qquad j = 1,\ldots,J,
\end{equation}
where $\textbf{U} \subseteq \textbf{Z}$, $\textbf{B}(t)$ is an $r$-vector of pre-specified time functions and $\mathbf{\eta}_j = (\mathbf{\kappa}_j, \mathbf{\upsilon}_j)$ stands for an $m$-vector of unknown regression parameters, $j = 1,\ldots,J$. For identifiability, we set $\mathbf{\eta}_J \equiv 0$. Examples of $\textbf{B}(t) $ are polynomial or spline functions. Denote by $\btheta$ the vector $(\bbeta, \bgamma, \mathbf{\eta}_1,\ldots, \mathbf{\eta}_{J})$ of all regression parameters characterizing the components of VMCF.

\subsection{Marginal model}

It is also interesting to look at the relationship between the marginal (population) total hazard $\lambda_{\bullet,\textrm{pop}}(t)$ and the conditional (on $Y=1$) total hazard $\lambda_{\bullet}(t|Y=1)$. First, note that the conditional expectation of $Y$ given $T\geq t$ is determined by
\begin{equation}\label{eq:cond_aver_of_Y}
  E[Y|T\geq t] = P(Y=1|T\geq t) = \frac{P(Y=1) P(T\geq t|Y=1)}{P(T\geq t)}=\frac{p \cdot S(t|Y=1)}{p  S(t|Y=1) + 1 - p}.
\end{equation}

The marginal total hazard is given by
\begin{equation}\label{eq:pop_haz}
  \lambda_{\bullet,\textrm{pop}}(t) = \frac{-S_{pop}^{'}(t)}{S_{pop}(t)} = \frac{p S(t|Y=1)\lambda_{\bullet}(t|Y=1) }{p  S(t|Y=1) + 1 - p}= E[Y|T\geq t] \lambda_{\bullet}(t|Y=1),
\end{equation}
where $S^{'}$ stands for the derivative of $S$. Intuitively, the above relationship between the two hazards expresses the fact that $\lambda_{\bullet,\textrm{pop}}(t)$ is the average of the $Y\lambda_{\bullet}(t|Y=1)$ taken over all the individuals at risk just before $T=t$.

Note that (\ref{eq:cond_haz_pop_model}) implies that the marginal total hazard is given by
\begin{equation}\label{eq:haz_pop_model}
   \lambda_{\bullet,\textrm{pop}}(t|\textbf{Z},\textbf{X}) = \lambda_{0}(t|Y=1 )\exp{(\bgamma^{\top} \textbf{Z})}
   \frac{g^{-1}(\bbeta^{\top} \textbf{X}^{*})S_0(t|Y=1)^{\exp{(\bgamma^{\top} \textbf{Z})}}}{g^{-1}(\bbeta^{\top} \textbf{X}^{*})S_0(t|Y=1)^{\exp{(\bgamma^{\top} \textbf{Z})}} + 1 - g^{-1}(\bbeta^{\top} \textbf{X}^{*})}\ ,
\end{equation}
where $S_0(t|Y=1) = \exp(-\Lambda_{0}(t|Y=1))$. This clearly shows that at the population level, the proportional hazards assumption postulated in the strata $\{Y=1\}$ no longer holds. Alternatively, we can write the model in the following way:
\begin{equation}\label{eq:haz_pop_its_model}
   \lambda_{\bullet,\textrm{pop}}(t|\textbf{Z}) = \lambda_{0}(t)\exp{(\mathbf{\eta}^{\top}(t) \textbf{Z})}\ ,
\end{equation}
where $\lambda_{0}(t)$ is an unspecified baseline hazard and $\mathbf{\eta}(t)$ is an unknown, time-dependent vector of regression coefficients. For two individuals with covariate vectors $\textbf{Z}$ and $\widetilde{\textbf{Z}}$, the following relationship holds for the $k$-th component of the vector $\mathbf{\eta}(t)$:
\begin{equation}\label{eq:log_haz_ratio}
    \mathbf{\eta}_{k}(t) (Z_{k}-\widetilde{Z}_{k}) = \mathbf{\gamma}_{k} (Z_{k}-\widetilde{Z}_{k}) + \log \frac{E[Y|T\geq t, \textbf{Z}, \textbf{X}]}{E[Y|T\geq t, \widetilde{\textbf{Z}},\widetilde{\textbf{X}}]}\ ,
\end{equation}
which expresses that the log hazard ratio at the population level equals the sum of the log hazard ratio in the strata $\{Y=1\}$  and of a time-varying term. Derivation of this formula is given in Appendix $A$.

\subsection{Likelihood}

Omitting covariates for a moment, the observed likelihood is the product of contributions of individuals from two categories: an individual $i$ who fails at time $t_i$ due to cause $j$ contributes
\begin{equation*}
    P(\widetilde{T}_i= t_i,D_i=j,Y_i = 1) = P(Y_i = 1) \cdot P(\tilde{T} = t_i| Y_i = 1) \cdot P(D_i = j| \tilde{T}_i = t_i) \cdot P(C_i > t_i),
\end{equation*}
and an individual $i$ who is censored at time $t_i$ contributes
\begin{equation*}
   P(\tilde{T} > t_i) = \big[P(Y_i = 1) \cdot P(\tilde{T} > t_i| Y_i = 1) + P(Y_i = 0)\big]\cdot P(C_i = t_i).
\end{equation*}

If we assume that the distributions of $T$ and $C$ have no common parameters, then we can omit the contribution of $C$ to the likelihood; therefore, the observed likelihood is proportional to
\begin{eqnarray*} L &=&
     \prod_{i=1}^{n} \Big[p_i P(\tilde{T} = t_i| Y_i = 1)
     \prod_{j=1}^{J} P(D_i = j| \tilde{T}_i = t_i)^{\textbf{1}\{D_i = j\}}\Big]^{\textbf{1}\{D_i > 0\}}\\
     &\cdot& \prod_{i=1}^{n} \big[p_i P(\tilde{T} > t_i| Y_i = 1) + (1 - p_i)\big]^{\textbf{1}\{D_i = 0\}}.\notag
\end{eqnarray*}
In terms of relative and (conditional) total hazards, the observed likelihood can be written as
\begin{eqnarray*} L &=&
    \prod_{i=1}^{n} \Big\{p_i \lambda_{\bullet}(t_i|Y_i=1) \exp{[-\Lambda_{\bullet}(t_i|Y_i=1)]}
    \prod_{j=1}^{J} \pi_{j}(t_i)^{\textbf{1}\{D_i = j\}}\Big\}^{\textbf{1}\{D_i > 0\}  }\\
    &\cdot& \prod_{i=1}^{n} \big\{p_i \exp{[-\Lambda_{\bullet}(t_i|Y_i=1)]} + (1 - p_i)\big\}^{\textbf{1}\{D_i = 0\}}\notag
\end{eqnarray*}
which can be separated in the following way:
\begin{equation}\label{eq:factorisation_full_lik}
   L(\btheta, \lambda_{0}(t|Y=1)) = L_{1}(\bbeta, \bgamma, \lambda_{0}(t|Y=1); Y)\cdot L_{2}(\mathbf{\eta}_1,\ldots, \mathbf{\eta}_{J}),
\end{equation}
where
\begin{eqnarray*}
  L_{1}(\bbeta, \bgamma, \lambda_{0}(t|Y=1); Y) &=& \prod_{i=1}^{n} \big\{ p_i \lambda_{\bullet}(t_i|Y_i=1)
    \exp{[-\Lambda_{\bullet}(t_i|Y_i=1)]}\big\}^{\textbf{1}\{D_i > 0\}}\notag\\
    &\cdot& \prod_{i=1}^{n} \big\{p_i \exp{[-\Lambda_{\bullet}(t_i| Y_i=1)]} + (1 - p_i)\big\}^{\textbf{1}\{D_i = 0\}}\notag
\end{eqnarray*}
and
\begin{equation*}
 L_{2}(\mathbf{\eta}_1,\ldots, \mathbf{\eta}_{J}) = \prod_{i=1}^{n} \prod_{j=1}^{J} \pi_{j}(t_i)^{\textbf{1}\{D_i = j\}}.
\end{equation*}

\subsection{Estimation}

We will use a variant of the EM algorithm given in \cite{SyTay:00} to estimate the parameters $(\btheta, \lambda_{0}(t_{1}|Y=1), \ldots, \lambda_{0}(t_{K}|Y=1))$ of the VMCF. The technique consists of an adaptation of the EM algorithm to deal with the latent $Y$ and to accommodate our semi-parametric approach. Typically, it is assumed that all event-free survivors after $t_{K}$ get cured. This amounts to imposing the zero-tail constraint, that is, $S(t|Y=1)=0$ for $t\geq t_{K}$, a condition which assures that the observed likelihood is well-behaved (see also \cite{SyTay:00}).

Denote by $\mathcal{C}$ the complete data, that is, the sample data when the latent $Y$ would be observed for all cases. The complete likelihood function is given by:
\begin{equation*}
   L_{\mathcal{C}}(\btheta, \lambda_{0}(t|Y=1)) = L_{3}(\bbeta, \bgamma, \lambda_{0}(t|Y=1); Y)\cdot L_{2}(\mathbf{\eta}_1,\ldots, \mathbf{\eta}_{J}),
\end{equation*}
where
\begin{eqnarray*}
  L_{3}(\bbeta, \bgamma, \lambda_{0}(t|Y=1); Y) &=& \prod_{i=1}^{n} \big\{ p_i \lambda_{\bullet}(t_i|Y_i=1)
    \exp{[-\Lambda_{\bullet}(t_i|Y_i=1)]}\big\}^{\textbf{1}\{D_i > 0\}}\notag\\
    &\cdot& \prod_{i=1}^{n} \Big\{ \big\{p_i \exp{[-\Lambda_{\bullet}(t_i| Y_i=1)]}\big\}^{\textbf{1}\{Y_i = 1\}} \cdot (1 - p_i)^{\textbf{1}\{Y_i = 0\}}\Big\}^{\textbf{1}\{D_i = 0\}}.\notag
\end{eqnarray*}
Note that $D_i > 0$ implies $Y_i = 1$, and $Y_i = 0$ implies $D_i = 0$. After rearranging the factors, we get:
\begin{eqnarray} L_{3}(\bbeta, \bgamma, \lambda_{0}(t|Y=1); Y) &=&
    \prod_{i=1}^{n} p_i^{\textbf{1}\{Y_i = 1\}} (1 - p_i)^{\textbf{1}\{Y_i = 0\}}\notag\\
    &\cdot& \prod_{i=1}^{n} \lambda_{\bullet}(t_i|Y_i=1)^{\textbf{1}\{D_i > 0\}}
    \exp{[-Y_i\Lambda_{\bullet}(t_i|Y_i=1)]} \label{eq:compl-like_vm} \\
    &=& L_{31}(\bbeta; Y) \cdot L_{32}(\bgamma, \lambda_{0}(t|Y=1); Y). \notag
\end{eqnarray}

The maximization of $\log L_{\mathcal{C}}(\btheta, \lambda_{0}(t|Y=1))$ is enhanced by the observation that the multinomial logistic structure embedded in $\log L_{2}(\mathbf{\eta}_1,\ldots, \mathbf{\eta}_{J})$ can be maximized independently of $\log L_{3}(\bbeta, \bgamma, \lambda_{0}(t|Y=1); Y)$. An appealing fact is that this can be achieved by means of standard software like the PROC GLM in \textsf{SAS} or the \textbf{\texttt{glm}} function in \textsf{R}~(\cite{R:10}). Denote by $(\widehat{\mathbf{\eta}}_1,\ldots, \widehat{\mathbf{\eta}}_{J})$
the estimator of $(\mathbf{\eta}_1,\ldots, \mathbf{\eta}_{J})$. Maximization of $\log L_{3}(\bbeta, \bgamma, \lambda_{0}(t|Y=1); Y)$ is more involved and uses the EM-algorithm following the approach in \cite{SyTay:00}. Technical details are given in Appendix $B$.

\subsection{Standard errors}

An approximation of the asymptotic variance of $(\btheta, \lambda_{0}(t_{1}|Y=1), \ldots, \lambda_{0}(t_{K}|Y=1))$ can be obtained as the inverse of the observed full information matrix $\Im_{\btheta, \lambda_{0}(t|Y=1)}$ of $L(\mathbf{\theta}, \lambda_{0}(t|Y=1))$. Note that due to the factorisation (\ref{eq:factorisation_full_lik}), we can write:
\begin{equation*}\Im_{\btheta, \lambda_{0}(t|Y=1)}=
\left(
 \begin{array}{ccc}
    \Im_{\bbeta, \bgamma, \lambda_{0}(t|Y=1)}  & | & 0  \\
    -            & | & -                                \\
    0            & | & \Im_{\eeta}  \\
  \end{array}
\right)\,
\end{equation*}
where $\Im_{\bbeta, \bgamma, \lambda_\bullet(t|Y=1)}$ is the observed full information matrix of $L_{1}(\bbeta, \bgamma, \lambda_{0}(t|Y=1); Y)$ and $\Im_{\eeta}$ is the observed full information matrix of $L_{2}(\eeta_1,\ldots, \eeta_{J})$.

\cite{SyTay:00} derived a formula for $\Im_{\bbeta, \bgamma, \lambda_\bullet(\cdot|Y=1)}$. In Appendix $C$, we use their results to obtain the standard errors of the conditional (on $Y=1$) cause-specific cumulative hazards from vertical modeling with a cure fraction.

\section{Simulations}\label{sec:sim}

In this simulation study, we assess the performance of our estimators in the semi-parametric setting when the assumed model is correct. We simulate data for $n = 500$ individuals, each of whom can fail due to the disease (cause 1) or due to a competing cause (cause 2), or can be subjected to right-censoring. Assume a non-negligible proportion of individuals get cured. Individuals are followed over a period of at most $15$ years; random right-censoring, which is independent of survival time, occurred uniformly in the interval $[7,15]$ years. Assume $Z \sim N(0,1)$ is a continuous, normally distributed baseline covariate. We generate data as random samples drawn from a VMCF model (the ``true" model), where we assume that the cure indicator $Y$ follows a logistic regression model:
\begin{equation*}
      \textrm{logit}(P(Y=1|Z)) = \beta_0 + \beta_1 \cdot Z,
\end{equation*}
for $(\beta_0,\beta_1) \in \{(-0.62,1.24),(-1.38,0),(1.38,0)\}$ leading to three scenarios with various amounts of cure proportions, that is, $65\%$ for $Z=1$, $80\%$ and $20\%$ respectively. The latency part is specified by
\begin{equation*}
    \lambda_{\bullet}(t|Y=1, Z) = \lambda_{0}(t|Y=1)\exp{(\gamma Z)},
\end{equation*}
where $\gamma = 0.3$ and the baseline hazard $\lambda_{0}(t|Y=1)$ is constant, equal to $0.4$. The relative hazards are constant such that $\pi_1(t) = 0.25$ and $\pi_2(t) = 0.75$, favoring cause $2$ over cause $1$.

VMCF and VM are fitted to the simulated data. The incidence component of VMCF and the conditional (on $Y=1$) total hazard are assumed to have the same form as in the true model. The relative hazards are assumed constant over time, with no covariate effect; the observed proportions of cause specific events relative to the total number of events were employed as their estimates. The covariate $Z$ was therefore included in the two mixture components (incidence and latency). The probabilities $\widehat{F}_j(t|Y_i=1, Z_i)$ and $\widehat{F}_j(t|Z_i)$, $j=1,2$, are reported at each of the prediction time points $\{1,2,5,7,11\}$ for an individual $i$ with $Z_i=1$. In VM, a Cox proportional hazards model with $Z$ as predictor with unspecified baseline hazard is postulated for the total hazard. The relative hazards are identical with those in VMCF. The probabilities $\widehat{F}_j(t|Z_i)$, $j=1,2$, are reported at each of the prediction time points $\{1,2,5,7,11\}$ for an individual $i$ with $Z_i=1$.

We reported in Tables~\ref{tab:simulation1} and~\ref{tab:simulation2} the estimated bias and root mean squared error (RMSE) obtained for our estimators under various proportions of susceptibles. Each scenario was run $10 000$ times.

\begin{table}[h!]
\caption{\label{tab:simulation1}Estimated bias (root mean squared error) $\widehat{F}_j(t|Y=1, Z=1)$ with respect to the true cumulative incidence function $F_j(t|Y=1, Z=1)$, $j=1,2$, for the three scenarios.}
\begin{center}
\resizebox{17cm}{!}{
\begin{tabular}{c|cc|cc|cc}
   \hline\hline
	 &  \multicolumn{2}{c}{$65\%$ cure in $\{Z=1\}$} &  \multicolumn{2}{c}{$80\%$ cure}  &  \multicolumn{2}{c}{$20\%$ cure} \\
	 \hline
   t & $\widehat{F}_1(t|Y=1, Z=1)$ & $\widehat{F}_2(t|Y=1, Z=1)$ & $\widehat{F}_1(t|Y=1, Z=1)$ & $\widehat{F}_2(t|Y=1, Z=1)$ & $\widehat{F}_1(t|Y=1, Z=1)$ & $\widehat{F}_2(t|Y=1, Z=1)$ \\[0.1em]
    \hline\hline
    1
    &  -0.0001 (0.0162) &  -0.0005 (0.0314) &  -0.0003 (0.0239) &  -0.0008 (0.0500) &  -0.0004 (0.0119)  &  -0.0012 (0.0247) \\
    2
    & \ 0.0001 (0.0229) & \ 0.0003 (0.0350) & \ 0.0000 (0.0328) & \ 0.0002 (0.0559) &  -0.0002 (0.0164)  &  -0.0006 (0.0276) \\
    5
    & \ 0.0012 (0.0302) & \ 0.0036 (0.0337) & \ 0.0018 (0.0419) & \ 0.0057 (0.0495) & \ 0.0004 (0.0208)  & \ 0.0015 (0.0240) \\
    7
    & \ 0.0017 (0.0313) & \ 0.0052 (0.0335) & \ 0.0030 (0.0435) & \ 0.0092 (0.0472) & \ 0.0008 (0.0215)  & \ 0.0027 (0.0228) \\
    10
    & \ 0.0029 (0.0321) & \ 0.0088 (0.0339) & \ 0.0049 (0.0447) & \ 0.0151 (0.0476) & \ 0.0013 (0.0219)  & \ 0.0041 (0.0225) \\
    \hline\hline
\end{tabular} }
\end{center}
\end{table}

\begin{table}[h!]
\caption{\label{tab:simulation2}Estimated bias (root mean squared error) of $\widehat{F}_j(t|Z=1)$ with respect to the true cumulative incidence function $F_j(t|Z=1)$, $j=1,2$, in VM and VMCF respectively, when $\beta_0 = -0.62$, $\beta_1 = 1.24$ and $\gamma = 0.3$.}
\begin{center}
{\small{
\begin{tabular}{c|cc|cc}
    \hline\hline
      &          \multicolumn{2}{c}{VM}    & \multicolumn{2}{c}{VMCF} \\
		\hline	
    t & $\widehat{F}_1(t|Z=1)$ & $\widehat{F}_2(t|Z=1)$ & $\widehat{F}_1(t|Z=1)$ & $\widehat{F}_2(t|Z=1)$ \\[0.1em]
    \hline\hline
    1
    & -0.0064 (0.0119) & -0.0193 (0.0284) & \ 0.0000 (0.0112)  & \ 0.0002 (0.0229) \\
    2
    & -0.0086 (0.0170) & -0.0260 (0.0366) & \ 0.0004 (0.0160)  & \ 0.0012 (0.0281) \\
    5
    & -0.0084 (0.0217) & -0.0254 (0.0393) & \ 0.0012 (0.0214)  & \ 0.0037 (0.0322) \\
    7
    & -0.0071 (0.0221) & -0.0214 (0.0376) & \ 0.0016 (0.0222)  & \ 0.0049 (0.0330) \\
    10
    & -0.0053 (0.0222) &  -0.0159 (0.0353) & \ 0.0024 (0.0228) & \ 0.0073 (0.0338) \\
    \hline\hline
\end{tabular} }}
\end{center}
\end{table}

The proposed method of estimation VMCF shows little bias in estimating both $\widehat{F}_j(t|Y=1, Z=1)$ and $\widehat{F}_j(t|Z=1)$, with bias (RMSE) increasing with the increase in the cure proportion. The bias contributes little to the RMSE. The behaviour at later time points is more uncertain due to the smaller number of events by the end of the follow up. Comparisons of $\widehat{F}_j(t|Z=1)$ derived from VMCF and VM with the true cause-specific cumulative incidences consistently show smaller bias (RMSE) for VMCF, thus supporting the idea that fitting VMCF to data from a population with a non-neglijable proportion of cure can replace a conventional survival model when the interest is to estimate the disease frequency. 

The coverage rates for the normal approximations $95\%$ confidence intervals of $\widehat{F}_j(t|Y=1, Z=1)$ and $\widehat{F}_j(t|Z=1)$ are higher for VMCF compared to VM. They are reported in Table~\ref{tab:simulation3} for the first scenario. The other two scenarios, with heavy or small amount of cure, show the same pattern (not reported).

\begin{table}[h!]
\caption{\label{tab:simulation3}Estimated coverage probabilities of $\widehat{F}_j(t|Y=1, Z=1)$ and $\widehat{F}_j(t|Z=1)$, $j=1,2$ in VM and VMCF respectively, when $\beta_0 = -0.62$, $\beta_1 = 1.24$ and $\gamma = 0.3$.}
\begin{center}
\resizebox{16cm}{!}{
\begin{tabular}{c|cc|cc|cc}
    \hline\hline
      &             \multicolumn{2}{c}{VMCF}                       &       \multicolumn{2}{c}{VM}                      &      \multicolumn{2}{c}{VMCF}       \\
		\hline	
    t & $\widehat{F}_1(t|Y=1, Z=1)$ & $\widehat{F}_2(t|Y= 1, Z=1)$ & $\widehat{F}_1(t| Z=1)$ & $\widehat{F}_2(t| Z=1)$ & $\widehat{F}_1(t| Z=1)$ & $\widehat{F}_2(t| Z=1)$ \\[0.1em]
    \hline\hline
    1
    & 95.2 & 95.4 & 90.8 & 85.1 & 95.2 & 95.5 \\
    2
    & 95.1 & 95.1 & 91.6 & 82.9 & 95.0 & 95.0 \\
    5
    & 95.1 & 95.1 & 93.3 & 86.8 & 95.1 & 95.2 \\
    7
    & 95.2 & 95.0 & 94.2 & 89.9 & 95.1 & 94.8 \\
    10
    & 95.1 & 94.4 & 94.8 & 92.5 & 95.1 & 94.6 \\
    \hline\hline
\end{tabular} }
\end{center}
\end{table}

The performance of estimators $\widehat{\beta_0}$, $\widehat{\beta_1}$ and $\widehat{\gamma}$ was studied in~\cite{SyTay:00}. Since these parameters are separated from the parameters of the relative hazards model, the results obtained in~\cite{SyTay:00} are applicable to and were confirmed in our setting and consequently have not been explored further.

\section{Data analysis}\label{sec:data_an}

\begin{figure}
\begin{center}
  \includegraphics[width=0.45\textwidth]{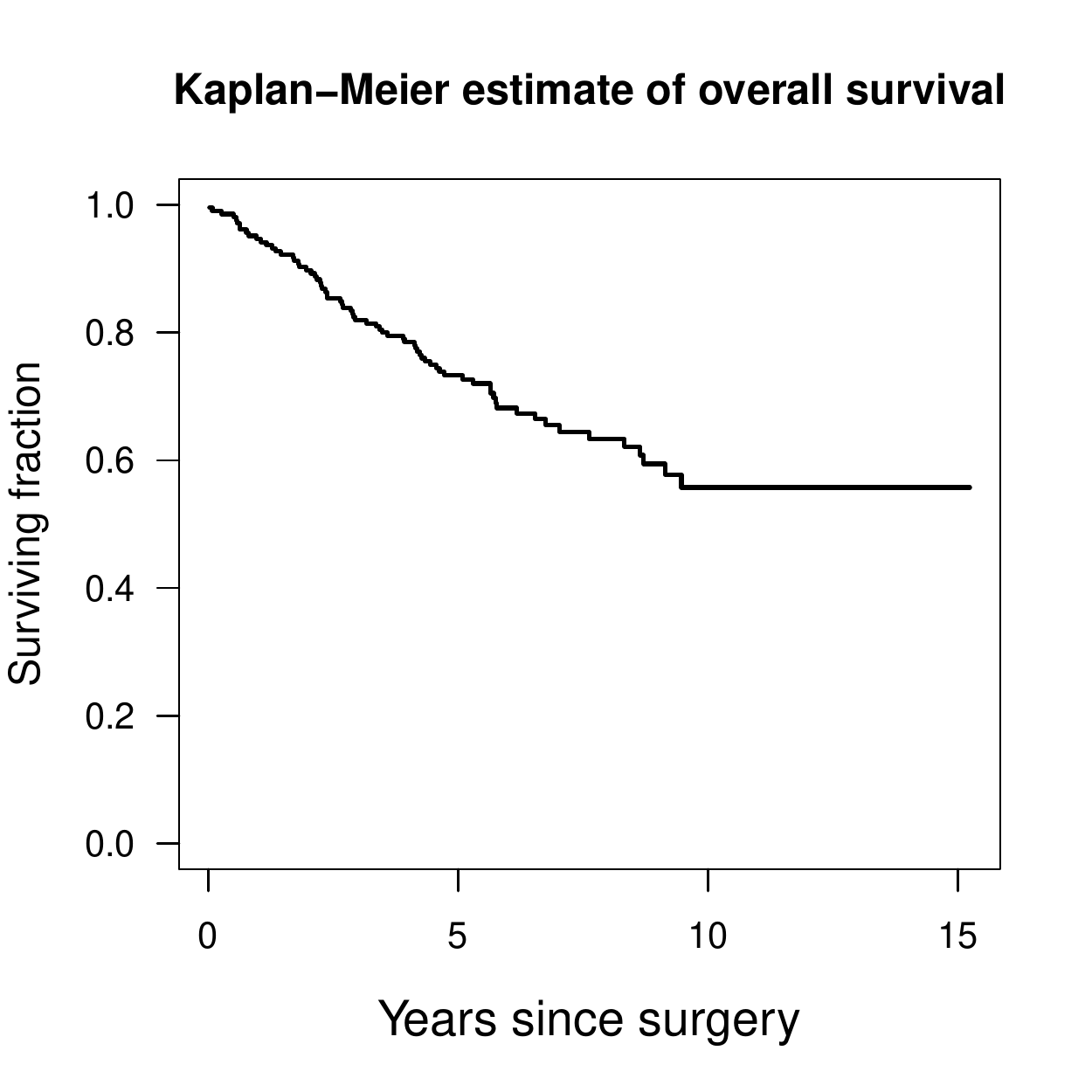}
  \caption{The estimated Kaplan-Meier curve of time to all-causes failure.}\label{fig:KM1}
\end{center}
\end{figure}

Our data comprises $205$ patients with malignant melanoma who had their tumor removed by surgery between 1962 and 1977. These data from~\cite{And:93} are part of the \textbf{\texttt{boot}} package~(\cite{Can:08}) for the \textsf{R} software. This was a prospective clinical study to assess the effect of risk factors on survival. Baseline covariates are gender, age (in years) at operation, year of operation, tumor thickness and ulceration of the tumor tissue. Age (scaled by 10), year of operation (centered at 1970 and scaled by 10) and thickness were considered continuous covariates; age ranged from $4$ to $95$ years and thickness ranged from $0.10$ to $17.42$ mm. $61\%$ of patients were men and $44\%$ of patients  presented with ulceration of the tissue at the time of surgery. The follow-up time ranged from $10$ days to $15.23$ years. The survival time is known only for those patients who have had their event before the end of 1977. The rest of the patients are censored at the end of 1977. Endpoint of interest is the time from operation to death due to melanoma (cause 1), in the presence of a competing cause, that is, death due to other causes, unrelated to melanoma (cause 2). The total number of deaths was $71$: $57$ $(80\%)$ deaths due to cause $1$ and $14$ $(20\%)$ deaths due to cause $2$. Of the $134$ censored patients, $27$ were censored with longer follow-up time than the largest death time.  Figure~\ref{fig:KM1} shows the Kaplan-Meier estimate of the survival function of time to any terminal event. This survival curve  appears to reach a plateau $10$ years post-surgery, indicating the presence of a sub-population which survives event-free by the end of the follow-up and clearly suggesting the appropriateness of a competing risks mixture cure model. For these data clinical interest is to understand how covariates affect the survival distribution, after surgery for melanoma, in the presence of competing risks.

First a vertical model ignoring cure (VM) was fit to the data. This model serves two purposes: (1) to estimate the effect of covariates on time to death due to each of  melanoma cancer and other causes and (2) to estimate the probabilities of each type of events since surgery. The model aggregates the covariate effects on latency and incidence in an intricate way (see, for instance, equation~(\ref{eq:haz_pop_model})). A Cox proportional hazards model was used for the total hazard including all covariates:
\begin{equation*}
    \lambda_{\bullet}(t|\textbf{Z}) = \lambda_{0}(t)\exp{(\bgamma^{\top} \textbf{Z})},
\end{equation*} where $\lambda_{0}(\cdot)$ is an unspecified baseline hazard and $\mathbf{\gamma}$ stands for a vector of unknown regression parameters, and a logistic regression model was used for the relative hazards including all covariates and piecewise constant time functions, with cut-off points at the $0.25,0.5,0.75$ quartiles of the failure time distribution leading to:
\begin{equation}\label{eq:log_reg_rel_haz}
     \textrm{logit}(\pi_1(t| \widetilde{T} = t, \textbf{Z})) = \mathbf{\kappa}_1^{\top} \textbf{B}(t) + \mathbf{\upsilon}_1^{\top} \textbf{Z},
\end{equation}
where $\textbf{B}(t) = (\textbf{1}\{\textrm{t} \in (0,1.76]\}, \textbf{1}\{\textrm{t} \in (1.76,2.90]\}, \textbf{1}\{\textrm{t} \in (2.90,4.67]\}, \textbf{1}\{\textrm{t} \in (4.67,15.23]\})$ and $\mathbf{\eta}_1 = (\mathbf{\kappa}_1, \mathbf{\upsilon}_1)$ stands for a vector of unknown regression parameters. The estimated regression parameters are reported in Table~\ref{tab:reg_param_nocure}. 

\begin{table}
\caption{\label{tab:reg_param_nocure}Regression parameters in VM.}
\centering
\resizebox{16cm}{!}{
\begin{tabular}{c|c|c}
    \hline\hline
     Covariate &  Regression parameters (SE)  &  Regression parameters (SE) \\
      &   in Cox model for survival&   in logistic  model  for relative hazard\\
    \hline\hline
     Thickness     &  0.09 (0.04) &  0.01 (0.12) \\
     Ulcer         &  0.98 (0.26) &  1.46 (0.86) \\
     Age           &  0.02 (0.08) & -0.04 (0.24) \\
     Year (stand)  & -0.88 (0.55) & -0.95 (1.70) \\
     Sex           &  0.42 (0.24) &  0.30 (0.72) \\
     \hfill $\textbf{1}\{\textrm{t} \in (   0, 1.76]\}$  &  & 1.93 (1.87) \\
     \hfill $\textbf{1}\{\textrm{t} \in (1.76, 2.90]\}$  &  & 4.24 (2.08) \\
     \hfill $\textbf{1}\{\textrm{t} \in (2.90, 4.67]\}$  &  & 3.84 (1.80) \\
     \hfill $\textbf{1}\{\textrm{t} \in (4.67,15.23]\}$  &  & 2.61 (1.67) \\
    \hline\hline
\end{tabular}}
\end{table}

For illustration, we present the results of the model fit for a male patient with ulceration and for average values of the continuous covariates. Table~\ref{tab:reg_haz} gives the estimated relative hazards implied by fitting model~(\ref{eq:log_reg_rel_haz}) with associated standard errors. It can be seen that the dominating cause of death post surgery is melanoma cancer. 

\begin{table}
\caption{\label{tab:reg_haz}Estimated piecewise constant relative hazards of cause 1 and their standard error for a male patient with ulceration and for average values of the continuous covariates derived by fitting VM.}
\centering
\resizebox{14cm}{!}{
\begin{tabular}{c|cccc}
    \hline\hline
        &  $(0, 1.76]$  &  $(1.76, 2.90]$  &  $(2.90, 4.67]$ & $(4.67,15.23]$ \\
    \hline
     Cause 1 &  0.754 (0.141) &  0.968 (0.038) & 0.954 (0.044) & 0.858 (0.106) \\
    \hline\hline
\end{tabular}}
\end{table}

\begin{figure}
\begin{center}
  \includegraphics[width=0.45\textwidth]{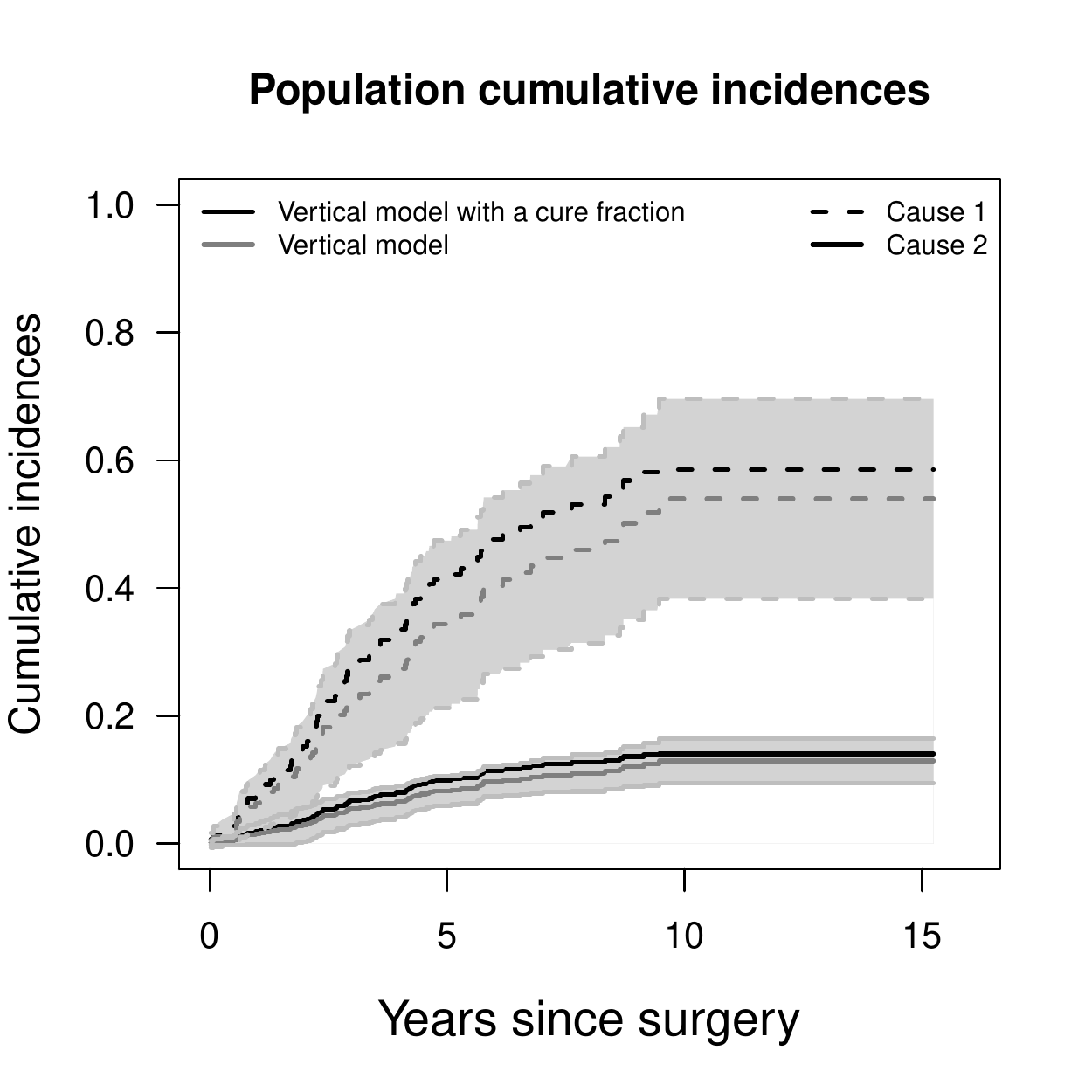}
  \caption{The estimated cumulative incidences of time to melanoma and time to other causes based on VM (in gray) and on VMCF (in black) for a male individual with ulceration and for the mean values of the continuous covariates, together with the $95\%$ confidence intervals derived by fitting the former model.}\label{fig:CumI}
\end{center}
\end{figure}

Figure~\ref{fig:CumI} shows (in gray lines) the estimated cumulative incidence functions of time to type 1-event and time to type 2-event.

We then applied the vertical modeling with a cured fraction (VMCF) to these data. This model serves three purposes: (1) to estimate the effect of covariates on time to death due to each of  melanoma cancer and other causes in the population of uncured patients, (2) to estimate the cause-specific cumulative incidences in the population of uncured patients and (3) to estimate the effect of covariates on the probability of being cured. A logistic regression model was postulated for the incidence part and for the latency part, a Cox proportional hazards model for the conditional (on $Y=1$) total hazard and a logistic model for the relative hazards where piece-wise constant time functions were used with cut-off points at the quartiles of the failure time distribution function. Note that this model for the relative hazards coincides with the one used in VM where the evidence of cure was ignored, because the conditional (on $T=t$) distribution of causes of failure depends only on the actual causes of failure observed. All baseline covariates were included in both parts (incidence and latency) of the model. No selection of covariates was performed to test whether some covariates could be removed. The estimated regression parameters are reported in Table~\ref{tab:reg_param}.

\begin{table}
\caption{\label{tab:reg_param}Regression parameters in VMCF.}
\centering
\begin{tabular}{c|c|cc}
    \hline\hline
                  &       \multicolumn{3}{c}{Regression parameters (SE)}   \\
    \hline\hline
    Covariate     &   Incidence model  &  \multicolumn{2}{|c}{Latency model}      \\
                  &                    &    Conditional Cox & Logistic      \\
    \hline
     Intercept    & -2.62 (0.88)  &                &               \\
     Thickness    &  0.07 (0.11)  &   0.10 (0.06)  &   0.01 (0.12) \\
     Ulcer        &  0.95 (0.64)  &   0.87 (0.50)  &   1.46 (0.86) \\
     Age          &  0.03 (0.15)  & -0.0006 (0.11)  &  -0.04 (0.24) \\
     Year(stand)  &  0.51 (0.89)  &  -1.49 (1.03)  &  -0.95 (1.70) \\
     Sex          &  0.57 (0.47)  &   0.42 (0.50)  &   0.30 (0.72) \\
  \hfill $\textbf{1}\{\textrm{t} \in (   0, 1.76]\}$  & &  & 1.93 (1.87) \\
  \hfill $\textbf{1}\{\textrm{t} \in (1.76, 2.90]\}$  & &  & 4.24 (2.08) \\
  \hfill $\textbf{1}\{\textrm{t} \in (2.90, 4.67]\}$  & &  & 3.84 (1.80) \\
  \hfill $\textbf{1}\{\textrm{t} \in (4.67,15.23]\}$  & &  & 2.61 (1.67) \\
  \hline\hline
\end{tabular}
\end{table}

\begin{figure}
\begin{center}
  \includegraphics[width=0.45\textwidth]{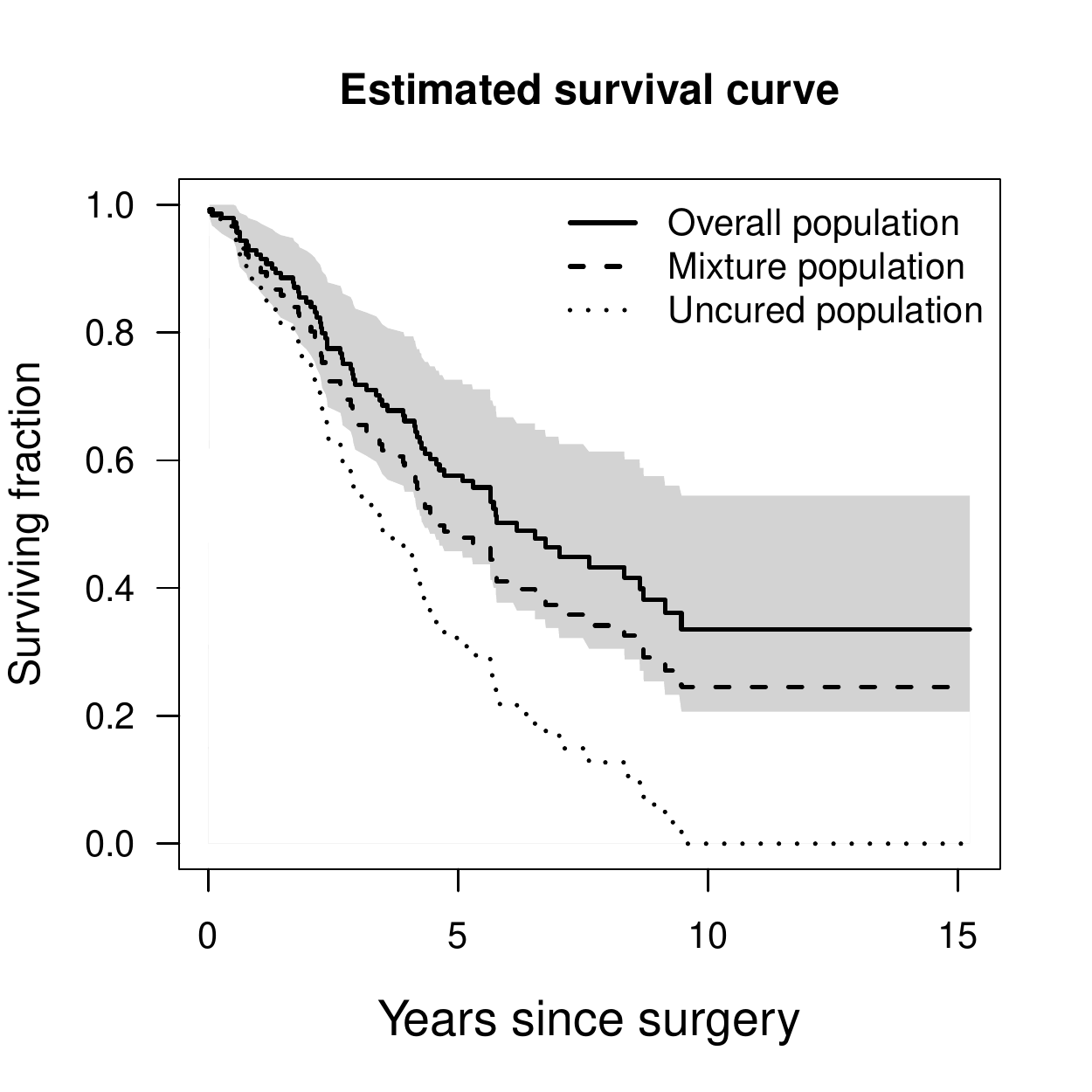}
  \caption{The estimated survival curve of time to any event at the mean values of the continuous covariates for a male individual with ulceration, based on VM (solid line) and on VMCF for the combined population (dashed line) and for the susceptible sub-population (dotted line), together with the $95\%$ confidence intervals derived by fitting the former model.}\label{fig:KM}
\end{center}
\end{figure}

The estimators of the survival curve $S_{\textrm{pop}}(t|\textbf{Z})$ and of the conditional (on $Y=1$) survival curve $S(t|Y=1, \textbf{Z})$, for a male individual with ulceration  and for average values of the continuous covariates,  are plotted in Figure~\ref{fig:KM} in dashed and dotted lines, respectively, and compared with the estimator of the survival curve $S(t|\textbf{Z})$ derived from VM (solid line).

\begin{figure}
\begin{center}
  \includegraphics[width=0.5\textwidth]{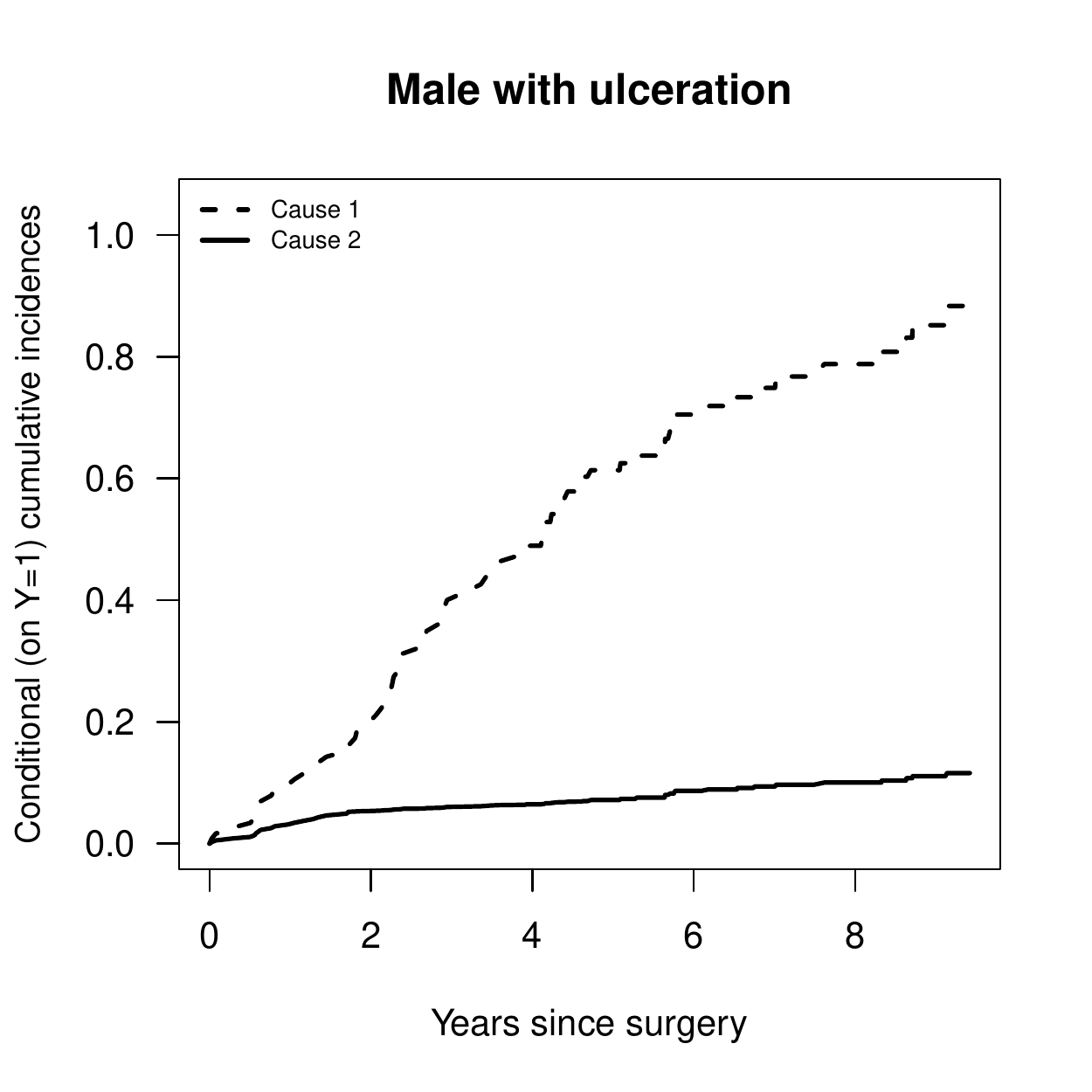}
  \caption{Model-based estimates of the conditional (on $Y=1$) cumulative incidences of time to melanoma and time to other causes for susceptible male patient with ulceration and for the mean values of the continuous covariates.}\label{fig:CumI_model_based}
\end{center}
\end{figure}

The estimated conditional (on $Y=1$) cumulative incidence function for a susceptible male patient with ulceration and for the mean values of the continuous covariates are plotted in Figure~\ref{fig:CumI_model_based}. The corresponding predicted probability of cure is $1-\widehat{p}=0.25$. We also plotted in Figure~\ref{fig:CumI} (in black) the estimated (unconditional) cumulative incidence functions of time to melanoma and time to other causes derived from VMCF, for a male individual with ulceration and average values of his other covariates. As expected in Figure~\ref{fig:KM}, the estimates for the population $S_{\textrm{pop}}(t|\textbf{Z})$ based on our VMCF model is similar to that  based on VM.

Finally, it is worth comparing Table~\ref{tab:reg_param_nocure} with Table~\ref{tab:reg_param}; it clearly shows that, especially for age, interpretation in terms of hazard ratios differs between $S(t|Y=1,\textbf{Z})$ and $S(t|\textbf{Z})$.


\section{Discussion}\label{sec:disc}

Little has been published about cure models in combination with competing risks. Several reviews have been recently published to highlight the advantages of cure models over the limitations of standard methods like Kaplan-Meier, Cox models or parametric models for survival data when statistical cure is a reasonable assumption~(\cite{Ji:13,Yu:13}). In this paper, we have illustrated a strategy for the statistical analysis of competing risks data with a cure fraction. In this way, we have argued that, when there is clinical evidence of a cured proportion in a cohort, special attention should be given to the heterogeneity present among individuals. 

An important advantage of the competing risks mixture cure model (VMCF) over the standard competing risks model (VM) is two-fold: (1) it allows inference of the susceptible sub-population, and therefore a better understanding and interpretation of the variability of the data and (2) it allows estimation and direct modeling of the cure indicator. Summary measures like these can be a useful tool to complement the existing statistical measures. Each covariate in VMCF can contribute with up to three sets of regression parameters, one parameter reflecting how the covariate affects the chance of cure, one parameter for the risk of failure, irrespective of the cause of failure, and one set of parameters for the relative position of each failure type among all failure types. 

An appealing technical feature of VMCF resides in its parametrisation, which is illustrated in equation~(\ref{eq:factorisation_full_lik}). VMCF naturally separates the observed likelihood into two factors: one where the cure indicator distribution is irrelevant and one where the cure indicator distribution is relevant. The former factor is pertinent to causes of failure when a failure occurs; in the absence of competing events, the observed likelihood~(\ref{eq:factorisation_full_lik}) reduces to the observed likelihood of the Cox proportional hazards mixture cure model of \cite{SyTay:00}. The latter factor uses information on the failure and censoring times and, therefore, it is sensitive to the joint distribution of $(T, Y)$. This feature makes our method straighforward to implement by means of \textrm{smcure} package available in~\cite{R:10}.

We believe that our approach can play a useful role, as it can accommodate complex competing risks data. It is worth mentioning that the approach can be naturally extended to deal with  missing causes of failure in competing risks. We refer to the work of~\cite{Nic:11} for a description of how this can be achieved. 

An important issue is that semi-parametric mixture cure models are by construction non-identifiable. We cannot precisely tell apart individuals who are cured or not among those who are censored. However, the presence of individuals with long follow-up and event-free works as empirical evidence of the existence of a cured subgroup. We adopt the strategy of \cite{SyTay:00}, who approach the non-identifiability problem through the use of the zero-tail constraint on the baseline failure time distribution. Another strategy has been adopted by \cite{Peng:03}, who impose a parametric shape on the tail of the failure time distribution.

Another aspect goes to the core of how cure is perceived in the presence of competing risks, other than the risk that is related to the disease. It might be the case that a diseased patient is at risk of several mutually exclusive causes of failure and their potential cure from the disease cannot be observed if they die from accidental causes. For instance, a different formulation is proposed by \cite{BasuTiw:10}, where separate competing risks structures are considered for the cure and the susceptible latent groups and a joint prior distribution is assumed on the collection of parameters. Cure is defined as not having experienced death due to the disease-related cause; therefore, the status of cure is observed only for individuals subjected to death not due to the cause of interest. We are currently adapting the vertical modeling approach to this situation.



\section*{Acknowledgements}

Research supported by IAP research network grant nr. P7/06 of the Belgian government (Belgian Science Policy). C. Legrand is supported by the contract 'Projet d'Actions de Recherche Concert\'ees' (ARC) 11/16-039 of the 'Communaut\'e francaise de Belgique', granted by the 'Acad\'emie universitaire Louvain'.

\bibliographystyle{chicago}
\bibliography{cure-cr}

\section*{Appendix A: Derivation of formula (\ref{eq:log_haz_ratio})}\label{sec:app_A}

Using (\ref{eq:haz_pop_its_model}), the hazard ratio corresponding to two individuals with covariate values given by $Z$ and $\widetilde{Z}$ is given by
\begin{equation*}
      \frac{\lambda_{\bullet,\textrm{pop}}(t|\textbf{Z})}{\lambda_{\bullet,\textrm{pop}}(t|\widetilde{\textbf{Z}})} =
      e^{\mathbf{\eta}^{\top}(t) (\textbf{Z}-\widetilde{\textbf{Z}})}.
\end{equation*}
On the other hand, using (\ref{eq:haz_pop_model}), the same hazard ratio is given by
\begin{equation*}
      \frac{\lambda_{\bullet,\textrm{pop}}(t|\textbf{Z}, \textbf{X})}{\lambda_{\bullet,\textrm{pop}}(t|\widetilde{\textbf{Z}},\widetilde{\textbf{X}})} =
      e^{\mathbf{\gamma}^{\top} (\textbf{Z}-\widetilde{\textbf{Z}})} \cdot \frac{E[Y|T\geq t, \textbf{Z}, \textbf{X}]}{E[Y|T\geq t, \widetilde{\textbf{Z}},\widetilde{\textbf{X}}]},
\end{equation*}
Equating the two representations leads to
\begin{equation*}
    \mathbf{\eta}^{\top}(t) (\textbf{Z}-\widetilde{\textbf{Z}}) = \mathbf{\gamma}^{\top} (\textbf{Z}-\widetilde{\textbf{Z}}) + \log \frac{E[Y|T\geq t, \textbf{Z}, \textbf{X}]}{E[Y|T\geq t, \widetilde{\textbf{Z}},\widetilde{\textbf{X}}]}
\end{equation*}
whose $k$-th component is given by (\ref{eq:log_haz_ratio}).

\section*{Appendix B: The EM-algorithm}\label{sec:app_B}

In the E-step of the algorithm the conditional expectation of $\log L_{3}(\bbeta, \bgamma, \lambda_{0}(t|Y=1); Y)$ is computed with respect to the distribution of the unobserved $Y_i$'s, given the current parameters values and the observed data $\mathcal{O} = (\mathcal{O}_i)_{i=1,\ldots,n}$. As $Y_i$'s contribute as linear terms in $\log L_{3}(\bbeta, \bgamma, \lambda_{0}(t|Y=1); Y)$, it is enough to compute, at a given iteration $m$, the weight $w_i^{(m)} = E(Y_i|\btheta^{(m)},\lambda^{(m)}_{0}(t|Y=1), \mathcal{O})$. For an individual $i$ who experiences an event at time $t_i$ (irrespective of its cause) the corresponding weight is
\begin{eqnarray*}
  w_i^{(m)} &=& E(Y|\btheta^{(m)},\lambda^{(m)}_{0}(t_i|Y=1), T = t_i, D_i \in\{1,\ldots,J\})\notag\\
  &=& P(Y_i=1|\btheta^{(m)},\lambda^{(m)}_{0}(t_i|Y=1), T = t_i, D_i \in\{1,\ldots,J\})\notag\\
  &=& 1,
\end{eqnarray*}
while for an individual  $i$ who is censored at time $t_i$ the corresponding weight is
\begin{eqnarray*}
  w_i^{(m)} &=& E(Y|\btheta^{(m)},\lambda^{(m)}_{0}(t_i|Y=1), T > t_i, D_i = 0)\notag\\
  &=& P(Y_i=1|\btheta^{(m)},\lambda^{(m)}_{0}(t_i|Y=1), T > t_i, D_i = 0)\notag\\
  &=& \frac{ g^{-1}(\bbeta^{\top} \textbf{X}_i^{*})\cdot S_{0}(t_i|Y=1)^{\exp(\bgamma^{\top} \textbf{Z}_i)}} { g^{-1}(\bbeta^{\top} \textbf{X}_i^{*})\cdot S_{0}(t_i|Y=1)^{\exp(\bgamma^{\top} \textbf{Z}_i)} + 1 - g^{-1}(\bbeta^{\top} \textbf{X}_i^{*})}\Biggr \rvert _{(\btheta, \lambda_{0}(t_i|Y=1)) = (\btheta^{(m)},\lambda^{(m)}_{0}(t_i|Y=1))}.
\end{eqnarray*}
Denote the expected complete log-likelihood by $\textbf{E}_{\textrm{p}}[\log L_{3}(\bbeta, \bgamma, \lambda_{0}(t_i|Y=1); w^{(m)})|\mathcal{O}]$, where $w^{(m)} = (w_i^{(m)})_{i=1,\ldots,n}$.

In the $M$-step of the algorithm, $\textbf{E}_{\textrm{p}}[\log L_{3}(\bbeta, \bgamma, \lambda_{0}(t|Y=1); w^{(m)})|\mathcal{O}]$ is maximized with respect to $(\bbeta, \bgamma, \lambda_{0}(t|Y=1)$, given $w^{(m)}$. Unlike in the standard Cox proportional hazards model, where the baseline hazard is seen as a nuisance parameter and eliminated in the procedure of estimating $\bgamma$, one cannot eliminate $\lambda_{0}(t|Y=1)$ in the Cox proportional hazards model embedded in the VMCF without loosing information about $\bbeta$. The main reason is that the hazard rate at the population level is no longer proportional (see (\ref{eq:haz_pop_model}) and (\ref{eq:haz_pop_its_model})) and arguments similar to those leading to the Cox partial likelihood do not hold true anymore. \cite{Peng:00}, \cite{SyTay:00} proposed a partial likelihood type method to estimate $\bgamma$ without specifying the nuisance parameter $\Lambda_{0}(t|Y=1)$. This involves the updating of the Aalen-Nelson estimator of $\Lambda_{0}(t|Y=1)$ at the $m$th iteration as given by
\begin{equation*}
      \widehat{\Lambda}_{0}(t|Y=1) = \sum_{i:t_{i}\leq t}\frac{d_i}{\sum_{l \in R_i}w_l^{(m)}\exp(\widehat{\bgamma}^{\top} \textbf{Z}_l)},
\end{equation*}
where $d_i$ is the number of events at time $t_i$, irrespective of the cause, and $R_i$ is the risk set at $t_i$. By substituting $\widehat{\Lambda}_{0}(t|Y=1)$ into $\log L_{32}(\bgamma, \lambda_{0}(t|Y=1); w^{(m)})$ we get the weighted partial likelihood of $\bgamma$, that is
\begin{equation*}
       \prod_{i=1}^{n}\Big[\frac{\exp(\bgamma^{\top} \textbf{Z}_i)}{\sum_{l \in R_i}w_l^{(m)}\exp(\bgamma^{\top} \textbf{Z}_l)}\Big]^{\textbf{1}\{D_i >0\}}.
\end{equation*}
To assure identifiability, we impose the zero-tail constraint (see \cite{SyTay:00}), that is, $\widehat{S}_0(t|Y = 1) = 0$ for $t > t_K$.

\section*{Appendix C: The standard errors of conditional (on $Y=1$) cumulative hazards from vertical modeling with a cure fraction}\label{sec:app_C}

In this appendix we derive the formula for the standard error of the conditional (on $Y=1$) cause-specific cumulative hazard from vertical modeling with a cure fraction, when we omit the vector $\textbf{Z}$ of covariates from the model of the latency component. As a consequence, the vector of parameter describing VMCF is $(\bbeta,\eeta,\lambda_{\bullet}(t_{1}|Y=1),\ldots, \lambda_{\bullet}(t_{K}|Y=1))$.

The Nelson-Aalen estimator of $\Lambda_\bullet(t|Y=1)$, denoted by $\widehat{\Lambda}_\bullet(t|Y=1)$, makes jumps of size $d\widehat{\Lambda}_\bullet(t|Y=1)$ at event time points $0 = t_0 \leq t_{1} < t_{2}<\ldots < t_{K}< \infty$. An approximation of the covariance matrix of $(\bbeta, d\widehat{\Lambda}_\bullet(t_{1}|Y=1),\ldots,d\widehat{\Lambda}_\bullet(t_{K}|Y=1))$ is given by the inverse of the observed full information matrix $\Im_{\bbeta, d\widehat{\Lambda}_\bullet(\cdot|Y=1)}$.

\textit{Remark}. The matrix $\Im_{\bbeta, d\widehat{\Lambda}_\bullet(\cdot|Y=1)}$ is not diagonal. Let $\XXi_{\bbeta, d\widehat{\Lambda}_\bullet(\cdot|Y=1)}$ denote an inverse of $\Im_{\bbeta, d\widehat{\Lambda}_\bullet(\cdot|Y=1)}$ and let $\XXi_{d\widehat{\Lambda}_\bullet(\cdot|Y=1)}$ be the sub-matrix of $\XXi_{\bbeta, d\widehat{\Lambda}_\bullet(\cdot|Y=1)}$ corresponding to $d\widehat{\Lambda}_\bullet(\cdot|Y=1)$.

Relevant quantities for our purposes are the relative hazards $\pi_j(t)$; we model them as in formula~(\ref{eq:rel_haz_model}).

\textit{Remark}. Often it will convenient to retain the system~(\ref{eq:rel_haz_model}) and to work with the $r\times J$ Fisher information matrix of
$\eeta^{\top} = \big(\eta_{1},\ldots,\eta_{J}\big)^{\top}$, denoted by $\Im_{\eeta}$ which has rank $r(J-1)$ and, in particular, is not invertible. Let $\XXi_{\eeta}$ denote a Moore-Penrose generalized inverse of $\Im_{\eeta}$.

We are interested to develop a formula for the $JK \times JK$ covariance matrix $\textrm{var}(\widehat{\LLambda}(\cdot|Y=1))=:\XXi_{\LLambda}$ of the estimator
$\widehat{\Lambda}_{j}(t|Y=1)=\sum_{t_s \leq t}\widehat{\pi}_j(t_s)\widehat{\lambda}_{\bullet}(t_s|Y=1)=\sum_{t_s \leq t}\widehat{\lambda}_{js, Y=1}$, where
$\widehat{\lambda}_{js, Y=1}=\widehat{\pi}_j(t_s)\widehat{\lambda}_{\bullet}(t_s|Y=1)$.
First, we introduce some notation, as follows:
\begin{equation*}
\btau = (\eeta,\lambda_{\bullet}(t_{1}|Y=1),\ldots, \lambda_{\bullet}(t_{K}|Y=1))^{\top}\ ,
\end{equation*}
\begin{equation*}
\LLambda=\big((\Lambda_j(t_{1}|Y=1))_{j=1,\ldots,J},(\Lambda_j(t_{2}|Y=1))_{j=1,\ldots,J},\ldots,(\Lambda_j(t_{K}|Y=1))_{j=1,\ldots,J} \big)^{\top}
\end{equation*}
and
\begin{equation*}
\llambda=\big(\lambda_{11,Y=1},\lambda_{21,Y=1},\ldots, \lambda_{J1,Y=1},\lambda_{12,Y=1},\ldots,\lambda_{J2,Y=1},\lambda_{1K,Y=1},\ldots,\lambda_{JK,Y=1} \big)\ .
\end{equation*}

According to the Delta-method, we get
\begin{equation}\label{eq:deltamethod}
\XXi_{\LLambda}=\frac{\partial\LLambda(\cdot|Y=1)}{\partial \llambda(\cdot|Y=1)}\cdot \frac{\partial\llambda(\cdot|Y=1)}{\partial\btau}\cdot \textrm{var}(\widehat{\btau})\cdot \Big(\frac{\partial\LLambda(\cdot|Y=1)}{\partial \llambda(\cdot|Y=1)}\cdot\frac{\partial\llambda(\cdot|Y=1)}{\partial\btau}\Big)^{\top}\ .
\end{equation}

It is straightforward to see that the matrix $\frac{\partial\LLambda}{\partial \llambda}$ of order $JK$ is given by
\begin{equation}\label{eq:jacobian cumulative hazard}
\frac{\partial\LLambda}{\partial \llambda}=
\left(
  \begin{array}{cccccc}
    I_{J\times J} & 0             & 0      &       &        & 0             \\
    I_{J\times J} & I_{J\times J} & 0      &       &        & 0             \\
                  &               & \ddots & \ddots&        &               \\
                  &               &        & \ddots& \ddots &               \\
                  &               &        &       & \ddots & 0             \\
    I_{J\times J} & I_{J\times J} &        &\ldots &        & I_{J\times J} \\
  \end{array}
\right)\,
\end{equation}
where $I_{J\times J}$ is the identity matrix of order $J$. Also, we have
\begin{equation*}
\frac{\partial \lambda_{js,Y=1}}{\partial \eta_{lu}}=-\pi_{j}(t_{s}) \pi_{l}(t_{s})\lambda_{\bullet}(t_{s}|Y=1)B_{u}(t_{s})\ ,
\end{equation*}
where $j,l \in \{1,\ldots,J\}$, $j \neq  l$, $s \in \{1,\ldots,K\}$,
$u \in \{1,\ldots, r\}$, and
\begin{equation*}
\frac{\partial \lambda_{js,Y=1}}{\partial \eta_{ju}}=\pi_{j}(t_{s})\big[1- \pi_{j}(t_{s})\big]\lambda_{\bullet}(t_{s}|Y=1)B_{u}(t_{s})\ ,
\end{equation*}
where $j\in \{1,\ldots,J\}$, $s \in \{1,\ldots,K\}$, $u \in \{1,\ldots, r\}$.

Moreover,
\begin{equation*}
\frac{\partial \lambda_{js,Y=1}}{\partial \lambda_{\bullet}(t_v|Y=1)}=\pi_{j}(t_{s}) \delta_{s,v}\ ,
\end{equation*}
where $j\in \{1,\ldots,J\}$, $s,v \in \{1,\ldots,K\}$ and $\delta$ stands for the Kronecker delta.

We shall define, for $t\geq 0$, the $J\times J$ matrix $\OOmega(t)$ as follows:
\begin{equation*}\OOmega(t)=
\left(
  \begin{array}{cccc}
    \pi_{1}(t)  & 0          & \ldots & 0          \\
    0           & \pi_{2}(t) & \ldots & 0          \\
    \ldots      & \ldots     & \ldots & \ldots     \\
    0           & 0          & \ldots & \pi_{J}(t) \\
  \end{array}
\right) -\big(\pi_{1}(t), \pi_{2}(t),\ldots,\pi_{J}(t)\big)^{\top}\big(\pi_{1}(t), \pi_{2}(t),\ldots,\pi_{J}(t)\big)\ ,
\end{equation*}
and the $r$-vector
\begin{equation*}
\aalpha(t)=\big(\lambda_{\bullet}(t|Y=1)B_{1}(t),\lambda_{\bullet}(t|Y=1)B_{2}(t),\ldots,\lambda_{\bullet}(t|Y=1)B_{r}(t) \big)^{\top}.
\end{equation*}
Setting
\begin{equation*}
\PPi(t_s)=\big(\pi_{1}(t_{s}), \pi_{2}(t_{s}),\ldots, \pi_{J}(t_{s})\big)^{\top},
\end{equation*}
a column vector of length $J$, for $s\in \{1,\ldots, K\}$, we get
\begin{equation}
\frac{\partial (\lambda_{1s,Y=1},\lambda_{2s,Y=1},\ldots,\lambda_{Js,Y=1})} {\partial (\eta_{1},\eta_{2},\ldots,\eta_{J})}=\OOmega(t_s)\otimes (\aalpha(t_s))^{\top},\ s\in \{1,\ldots, K\}\ ,
\end{equation}
where $\otimes$ stands for the Kronecker product, and finally
\begin{equation}\label{eq:jacobian relative hazard}
\frac{\partial \llambda}{\partial \btau}=
\left(
      \begin{array}{ccccccc}
      \OOmega(t_1)\otimes (\aalpha(t_1))^{\top} & \PPi(t_1) &  0         & 0       &          &         & 0         \\
      \OOmega(t_2)\otimes (\aalpha(t_2))^{\top} & 0         & \PPi(t_2)  & 0       &          &         & 0         \\
                                                &           &     0      & \ddots  & \ddots   &         &           \\
      \vdots                                    & \vdots    &            & \ddots  & \ddots   & \ddots  &           \\
                                                &           &            &         & \ddots   & \ddots  &   0       \\
      \OOmega(t_K)\otimes (\aalpha(t_K))^{\top} & 0         &  0         & 0       &  \ldots  &    0    & \PPi(t_K) \\
      \end{array}
      \right)
\end{equation}
which is a matrix of order $JK \times (Jr+K)$. As a result, we obtain
\begin{equation}\label{eq:covariance relative hazard}
\XXi_{\llambda}=\frac{\partial \llambda}{\partial \btau}
\left(
  \begin{array}{ccc}
    \XXi_{\eeta}  & | & 0                                             \\
    -              & | & -                                             \\
    0              & | & \XXi_{d\widehat{\Lambda}_\bullet(\cdot|Y=1)}  \\
  \end{array}
\right)\Big(\frac{\partial \llambda}{\partial \btau}\Big)^{\top}\ .
\end{equation}

In conclusion, using (\ref{eq:deltamethod}), (\ref{eq:jacobian cumulative hazard}) and (\ref{eq:covariance relative hazard}), we
have that
\begin{equation}
\XXi_{\LLambda}=\left(
                   \begin{array}{cccc}
                     \mathbf{W}_1 \XXi_{\eeta}\mathbf{W}_1+\widetilde{\PPi}_1 & \mathbf{W}_1 \XXi_{\eeta}\mathbf{W}_2+\widetilde{\PPi}_1 & \ldots & \mathbf{W}_1 \XXi_{\eeta}\mathbf{W}_K+\widetilde{\PPi}_1 \\
                     \mathbf{W}_2 \XXi_{\eeta}\mathbf{W}_1+\widetilde{\PPi}_1 & \mathbf{W}_2 \XXi_{\eeta}\mathbf{W}_2+\widetilde{\PPi}_2 & \ldots & \mathbf{W}_2 \XXi_{\eeta}\mathbf{W}_K+\widetilde{\PPi}_2 \\
                     \vdots                & \vdots                & \ddots & \vdots                \\
                     \mathbf{W}_K \XXi_{\eeta}\mathbf{W}_1+\widetilde{\PPi}_1 & \mathbf{W}_K \XXi_{\eeta}\mathbf{W}_2+\widetilde{\PPi}_2 & \ldots & \mathbf{W}_K \XXi_{\eeta}\mathbf{W}_K+\widetilde{\PPi}_K \\
                   \end{array}
                 \right)\ ,
\end{equation}
where
\begin{equation*}
\mathbf{W}_k=\sum_{s=1}^{k}\OOmega(t_s)\otimes (\aalpha(t_s))^{\top},\ k \in \{1,\ldots,K\}\ ,
\end{equation*}
and
\begin{equation*}
\widetilde{\PPi}_{kl}=\sum_{l=1}^{k}\XXi_{d\widehat{\Lambda}_\bullet(\cdot|Y=1)} \otimes \big(\sum_{s=1}^{k} \PPi(t_s)(\PPi(t_l))^{\top}\big),\ k,l \in \{1,\ldots,K\}\ .
\end{equation*}

\end{document}